\def\rn{\noindent\parshape 2 0truecm 8.8truecm 0.3truecm 8.5truecm}
\def\nn#1 #2{#1, #2.}				
\def\nnn#1 #2 #3{#1, #2. #3.}			
\def\nnnn#1 #2 #3 #4{#1, #2. #3. #4.}		
\def\nnnnn#1 #2 #3 #4 #5{#1, #2. #3. #4. #5.}	
\def\dualand{, \&\hbox{ }}				
\def\multiand{, \&\hbox{ }}				

\def\rg#1;#2;#3;#4;#5;#6 {\par\rn#1 #2, {\it #3}, {\bf #4}, #5 (``#6'') \par}
\def\rf#1;#2;#3;#4;#5 {\par\rn#1 #2, {\it #3}, {\bf #4}, #5\par}
\def\rfbook#1;#2;#3;#4;#5 {{\frenchspacing\par\rn#1 #2, {\it #3} (#4: #5)\par}}
\def\rfproc#1;#2;#3;#4;#5;#6 {{\frenchspacing\par\rn#1 #2, in {\it #3}, ed. #4 (#5: #6)\par}}
\def\rfprep#1;#2;#3  {{\par\rn#1 #2, #3\par}}
\def\rfprepp#1;#2;#3 {{\par\rn#1 #2, #3\par}}

\def\second{{\rm s}}

\def\Mpc{{\rm Mpc}}

\def\expec#1{\langle#1\rangle}

\def\etal{{\frenchspacing\it et al.}}
\def\ie{{\frenchspacing\it i.e.}}
\def\eg{{\frenchspacing\it e.g.}}
\def\etc{{\frenchspacing\it etc.}}
\def\rms{rms}

\def\beq#1{\begin{equation}\label{#1}}
\def\eeq{\end{equation}}
\def\beqa#1{\begin{eqnarray}\label{#1}}
\def\eeqa{\end{eqnarray}}
\def\eq#1{equation~(\ref{#1})}
\def\Eq#1{Equation~(\ref{#1})}
\def\eqn#1{~(\ref{#1})}

\def\ns{\vskip-0.2truecm}
\def\bs{\hskip-0.2truecm}

\def\spose#1{\hbox to 0pt{#1\hss}}
\def\simlt{\mathrel{\spose{\lower 3pt\hbox{$\mathchar"218$}}
     \raise 2.0pt\hbox{$\mathchar"13C$}}}
\def\simgt{\mathrel{\spose{\lower 3pt\hbox{$\mathchar"218$}}
     \raise 2.0pt\hbox{$\mathchar"13E$}}}
\def\simpropto{\mathrel{\spose{\lower 3pt\hbox{$\mathchar"218$}}
     \raise 2.0pt\hbox{$\propto$}}}

\def\ed{\end{document}}

\def\rt{\rho}
\def\rl{\rho_g}
\def\dt{\delta}
\def\dl{g}

\def\b{{\bf b}}
\def\g{{\bf g}}
\def\Dg{\Delta\g}
\def\e{{\bf e}}
\def\k{{\bf k}}
\def\r{{\bf r}}
\def\x{{\bf x}}
\def\nc{n_c}
\def\nf{n_f}
\def\ns{n_s}
\def\xh{\widehat{\x}}
\def\E{{\bf E}}
\def\F{{\bf F}}
\def\G{{\bf G}}
\def\I{{\bf I}}
\def\N{{\bf N}}
\def\P{{\bf P}}
\def\Q{{\bf Q}}
\def\R{{\bf R}}
\def\SS{{\bf S}}
\def\Px{P_\times}
\def\tr{\hbox{tr}\>}

\def\Nbar{\bar N}
\def\ith{i^{th}}
\def\first{1^{st}}
\def\second{2^{nd}}
\def\shot{\varepsilon}

\def\bbbcomment#1{}

\documentstyle[emulateapj]{article}
\begin{document}


\journalid{337}{15 January 1989}
\articleid{11}{14}

\submitted{Submitted to ApJL September 24, 1998; accepted April 13}

\vspace*{-0.5in}

\title{OBSERVATIONAL EVIDENCE FOR STOCHASTIC BIASING}

\author{
Max Tegmark$^{1,2}$
and
Benjamin C.~Bromley$^3$
}

\begin{abstract}
We show that the galaxy density in the Las Campanas Redshift Survey
(LCRS) cannot be perfectly correlated with the underlying mass
distribution since various galaxy subpopulations are not perfectly
correlated with each other, even taking shot noise into account.  This
rules out the hypothesis of simple linear biasing, and suggests that
the recently proposed stochastic biasing framework is necessary for
modeling actual data.
\end{abstract}




\section{INTRODUCTION}

Measurements of clustering in upcoming galaxy redshift surveys hold
the potential of measuring cosmological parameters with great accuracy
(Tegmark 1997; Goldberg \& Strauss 1997), especially when complemented
by measurements of the Cosmic Microwave Background (Eisenstein {\etal}
1998; Hu {\etal} 1998ab).  Since such measurements are only as
accurate as our understanding of biasing, there has been a recent
burst of work on the relation between the distribution of galaxies and
the underlying mass.

Dekel and Lahav (1998; Dekel 1997 \S 5.5; Lahav 1996 \S 3.1) have
proposed a robust framework termed ``stochastic biasing'', which drops
the assumption that the galaxy density $\rl(\r)$ is uniquely
determined by the matter density $\rho$.  Writing the corresponding
density fluctuations as $\dl\equiv\rl/\expec{\rl}-1$ and
$\dt\equiv\rt/\expec{\rt}-1$, $\dl$ is modelled as a function of $\dt$
plus a random term.  
It has been known since the outset (\eg, Dressler 1980) 
that any deterministic biasing relation $\dl=f(\dt)$ must be complicated, 
depending on galaxy type. Even allowing this, however,
deterministic biasing still implies that the
peaks and troughs of the two fields must coincide spatially,
which need not be the case.
Restricting attention to second moments, all the
information about both such stochasticity and nonlinearity (of
$f(\dt)$) can be contained in a single new function $r$ (Pen 1998;
Tegmark \& Peebles 1998, hereafter TP98).  Grouping the densities into
a two-dimensional vector
\beq{xDefEq}
\x\equiv\left({\dt\atop\dl}\right)
\eeq
and assuming nothing except translational invariance, its Fourier
transform $\xh(\k)\equiv\int e^{-i\k\cdot\r}\x(\r)d^3r$ obeys
\beq{MatrixPowerEq}
\expec{\xh(\k)\xh(\k')^\dagger}=(2\pi)^3\delta^D(\k-\k')
\left(\bs\begin{tabular}{cc}
$P(\k)$ 	&$\Px(\k)$\\
$\Px (\k)$	&$P_g(\k)$
\end{tabular}\bs\right)
\eeq
for some $2\times 2$ power spectrum matrix that we will denote
$\P(\k)$.  Here $P$ is the conventional power spectrum of the mass
distribution, $P_g$ is the power spectrum of the galaxies, and $\Px$
is the cross spectrum.  It is convenient to rewrite this covariance
matrix as
\beq{PdefEq}
\P(\k ) =
P(\k)\left(\bs\begin{tabular}{cc}
$1$ & $b(\k)r(\k)$ \\ $b(\k)r(\k)$ & $b(\k)^2$ \end{tabular}\bs\right)
\eeq
\linebreak

\smallskip

{\footnotesize

$^1$ Institute for Advanced Study, Princeton, NJ 08540
 
     
$^2$ Hubble Fellow; max@ias.edu

     
$^3$ Dept. of Physics, Univ. of Utah, Salt Lake City, UT 84112
}

\bigskip
\goodbreak

\noindent
where $b\equiv(P_g/P)^{1/2}$ is the bias factor (the ratio of luminous
and total fluctuations) and the new function
$r\equiv\Px/(P
P_g)^{1/2}$ is the dimensionless correlation coefficient between
galaxies and matter.  The special case $r = 1$ gives the simple
deterministic biasing relation $\dl = b\dt$, however, both $b$ and $r$
may generally depend on scale.

Since the function $r(k)$ is a cosmologically important quantity,
it has received much recent attention. 
Pen (1998) has shown how it can be measured
using redshift space distortions and nonlinear effects, Scherrer \&
Weinberg (1998) have computed $r$ for a number of theoretical
models, and Blanton {\etal} 1998 (hereafter B98) have 
estimated $b$ and $r$ from hydrodynamic simulations.
TP98 have computed the time-evolution of bias in the linear regime,
while Taruya {\etal} (1998ab) have generalized this result to the
perturbatively non-linear case (see also Mo \& White 1996; Mo {\etal}
1997; Matarrese {\etal} 1997; Bagla 1998; Catelan 1998ab; Colin
{\etal} 1998; Lemsen \& Sheth 1998; Moscardini {\etal} 1998; Porciani
1998; Wechsler {\etal} 1998).  Some of these numerical and theoretical
predictions have been borne out in observational data which
indicate very high bias values ($b\sim 4-6$) at $z\sim 3$, decreasing
rapidly with time (Giavalisco {\etal} 1998).

In light of all this activity,
it would be timely to observationally measure $r(k)$.
This is the topic of the present paper.
Unfortunately, measuring $r$ directly requires knowledge of
the true matter distribution $\dt$. Although this information in
principle may be obtained from {\eg} POTENT reconstruction from
peculiar velocity measurements (Dekel {\etal} 1990), the quasilinear
redshift-space method of Pen (1998) or gravitational lensing (Van
Waerbeke 1998), it will likely require better data than is presently
available since systematic errors in the estimate of $\dt$ masquerade
as $r<1$.
We therefore adopt an indirect approach, based on the following simple
idea. If two different types of galaxies are both perfectly
correlated with the matter, then they must also be perfectly
correlated with each other.  If we can demonstrate imperfect
correlations between two galaxy subsamples
then we will have shown that $r<1$ for at least one of them.
%
%
%
We show that $r<1$ at high statistical significance,
and place quantitative limits on $r$ in \S3.


\section{Ruling out simple biasing}

We base our analysis on galaxies in the Las Campanas Redshift Survey
(LCRS; Schectman {\etal} 1996) grouped into 6 types or ``clans''
according to the spectral classification scheme of Bromley {\etal} (1998ab), 
ranging from very early-type galaxies (clan 1) to
late-type objects (clan 6). To reduce shot noise we define 
our clan 4 as the combination of the original clans 4, 5 and 6. The
clustering properties of these clans vary in a systematic way,
revealing a progression in the relative bias factors $b$ -- see also 
Santiago \& Strauss (1992).
B98 predict that $r$ depends strongly on the formation epoch of a galaxy
population, so one might suspect that some of the LCRS clans have
$r<1$.


\centerline{{\vbox{\epsfxsize=8.5cm\epsfbox{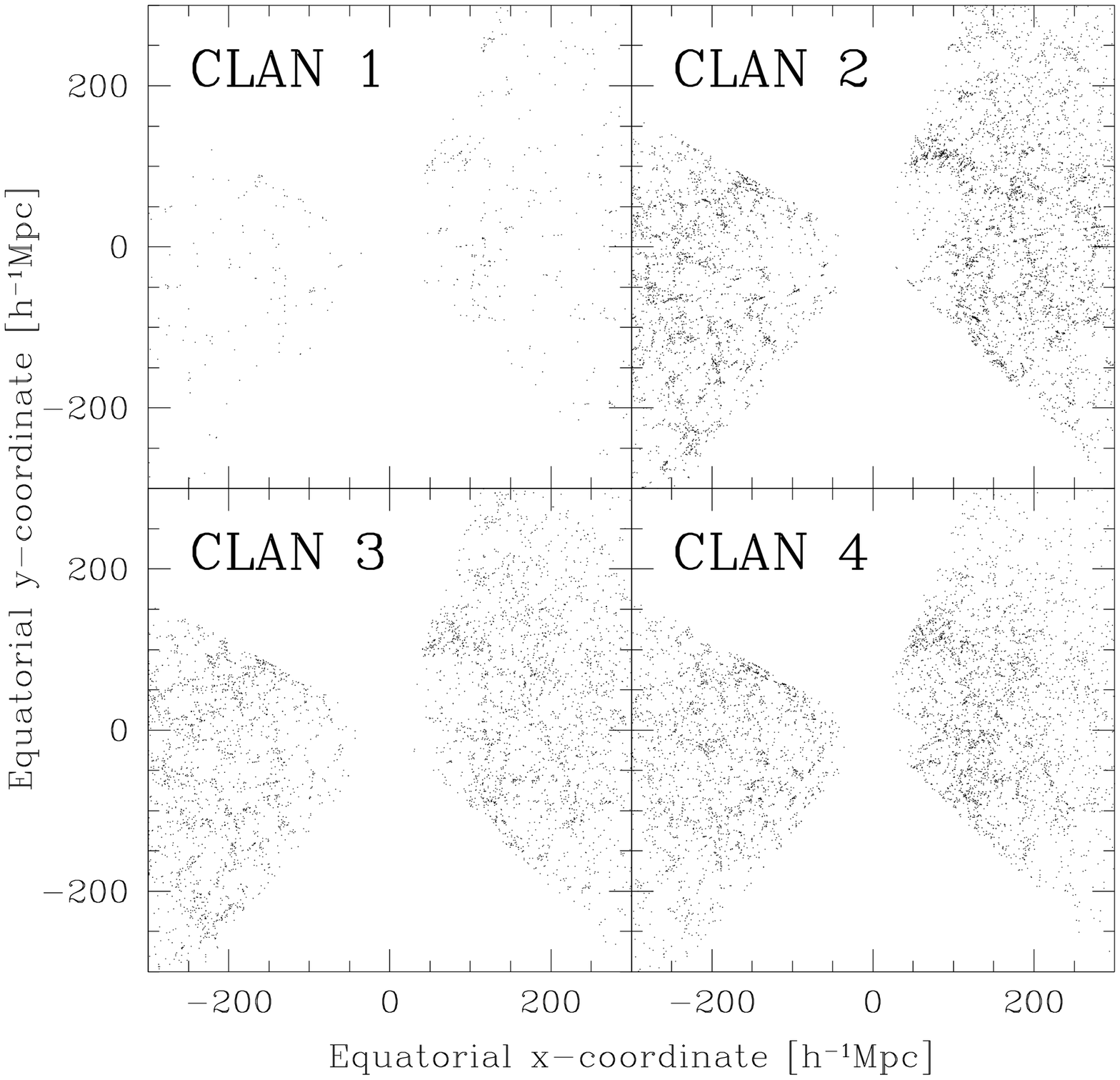}}}}
\vskip-0.3cm
{\footnotesize {\bf FIG. 1}
--- The four subsamples (clans) of the LCRS.
}



\bigskip
\noindent

The LCRS consists of $\nf=327$ rectangular fields in the sky, and we
further subdivide the volume into $\ns=10$ radial shells in the range
10 000 km/s$<cz<$45 000 km/s.  Discarding galaxies outside of this
redshift range (as in Lin {\etal} 1996) leaves 519, 8282, 5152 and
5669 galaxies in the four clans.  For each of our
$n\equiv\nf\times\ns$ spatial volumes $V_\alpha$ ($\alpha=1,...,n$)
and each clan $i=1,...,\nc$, we count the number of observed galaxies
$N^{(i)}_\alpha$, and compute the expected number of galaxies
$\Nbar^{(i)}_\alpha$ using the selection function of Lin {\etal}
(1996) with the clan-dependent Schechter parameters of Bromley {\etal}
(1998a).  We write the map of observed density fluctuations for the
$\ith$ clan as an $n$-dimensional vector $\g^{(i)}$, defined by
$g^{(i)}_\alpha\equiv N^{(i)}_\alpha/\Nbar^{(i)}_\alpha - 1$.  In a
simple $r=1$ linear biasing model, we would have
\beq{SimpleBiasEq}
g^{(i)}_\alpha = b_i \delta_\alpha + \shot^{(i)}_\alpha,
\eeq
where $b_i$ is the bias of the $\ith$ clan, $\delta_\alpha$ is the
matter density fluctuation in $V_\alpha$ and the shot noise
contributions $\shot^{(i)}_\alpha$ have zero mean and a diagonal
covariance matrix
\beq{NdefEq}
\N^{(i)}_{\alpha\beta}
\equiv\expec{\shot^{(i)}_\alpha \shot^{(i)}_\beta} 
= \delta_{\alpha\beta}/\Nbar^{(i)}_\alpha.
\eeq
Can we rule this out?  For a pair of clans $i$ and $j$, consider
the difference map
\beq{DifferenceEq}
\Dg\equiv\g^{(i)} - f\g^{(j)}.
\eeq
for different values of the factor $f$.  If $f=b_i/b_j$, then
\eq{SimpleBiasEq} shows that the (unknown) matter density fluctuations
$\delta_\alpha$ will cancel out, and $\Dg$
will consist of mere shot noise whose covariance matrix is
$\N\equiv\expec{\Dg\Dg^t}=\N^{(i)} + f^2 \N^{(j)}$.

Given the alternative hypothesis that there is a residual signal with
some covariance matrix $\SS$, so that $\expec{\Dg\Dg^t}=\N + \SS$, 
the most 
powerful ``null-buster'' test for ruling out the null hypothesis
$\expec{\Dg\Dg^t}=\N$ is using the generalized $\chi^2$-statistic
(Tegmark 1998)
\beq{NullbusterEq}
    \nu \equiv {\Dg^t\N^{-1}\SS\N^{-1}\Dg - \tr\N^{-1}\SS
    \over
    \left[2\tr\left\{\N^{-1}\SS\N^{-1}\SS\right\}\right]^{1/2}},
\eeq
which can be interpreted as the number of ``sigmas'' at which the
noise-only null hypothesis is ruled out.  We choose
$\SS_{\alpha\beta}=\xi(|\r_\alpha-\r_\beta|)$, where $\r_\alpha$ is
the center of volume $V_\alpha$ and $\xi$ is the correlation function
measured by the LCRS (Tucker {\etal} 1997).  We plot the results in
Figure 2 for 
three pairs
of clans $i$ and $j$, and the corresponding
valley-shaped curve tells us a number of things.  The fact that
$\nu\gg 1$ on the left-hand-side (as $f\to 0$, with all the weight on
clan $i$) means that there is a strong detection of cosmological
fluctuations above the shot noise level $(\nu\sim 1)$. Likewise,
$\nu\gg 1$ on the right-hand-side (as $f\to\infty$), which
demonstrates cosmological signal in clan $j$.  The fact that the curve
dips for intermediate $f$-values tells us that the two density maps
are correlated $(r>0)$ and have common signal. The minimum is attained
at the value $f$ which gives the best fit relative bias $b_i/b_j$ for
this common signal.  However, the fact that $\nu\gg 1$ even at the
minimum proves that even though some signal is shared in common, {\it not
all of it is}: there are no values of $b_i$ for which
\eq{SimpleBiasEq} can hold for any pair of clans $i$ and $j$.

Note that Figure 2 does not directly tell us which clans are most
correlated, since low minima can signal either strong correlations,
low fluctuations or high shot noise --- the latter being the case for
the (rare) $\first$ clan.

\smallskip
\centerline{{\vbox{\epsfxsize=8.8cm\epsfbox{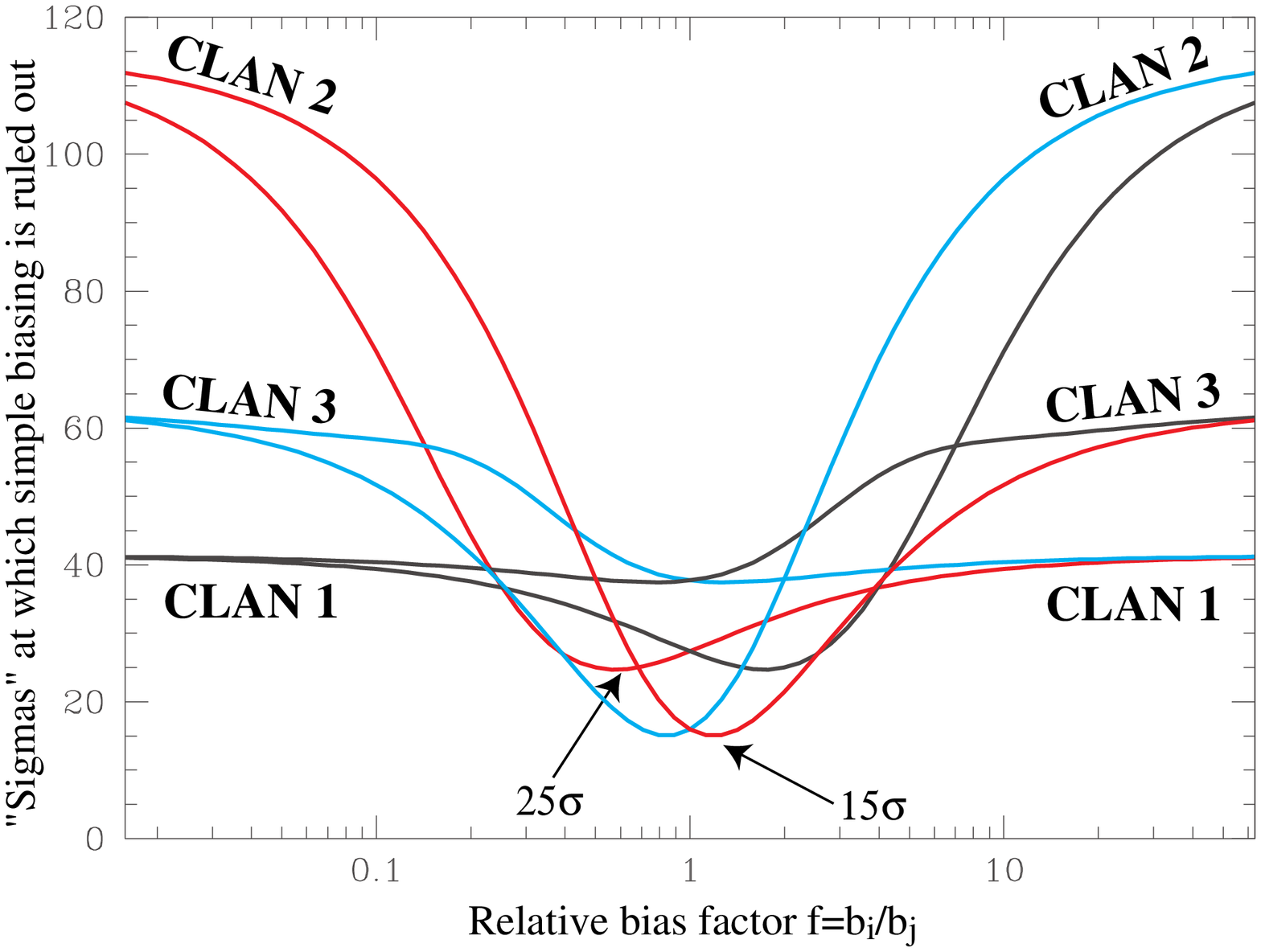}}}}
{\footnotesize {\bf FIG. 2} --- Significance at which we can rule out
that different clan pairs are perfectly correlated (that
$\g^{(i)}-f\g^{(j)}$ is pure shot noise).  }

\bigskip




\bigskip
\addtocounter{section}{1}
\centerline{\footnotesize \arabic{section}. MEASURING $r(k)$}
\smallskip

Having ruled out non-stochastic linear biasing, \ie, demonstrated that
$r<1$ for some clans, let us now study in more detail what constraints
can be placed on $r$.
%
\subsection{Upper limits on $r$}

Let us lengthen the
2-dimensional vector $\x=(\dt,\dl)$ from \eq{xDefEq} to an
$(1+\nc)$-dimensional vector including all $\nc$ galaxy clans
($\nc=4$): $x_0=\dt$ and $x_i=g^{(i)}$, $i=1,...,\nc$,
\ie,
$\x=(\dt,g^{(1)},g^{(2)},g^{(3)},g^{(4)})$.  Next we factor the
$(1+\nc)\times(1+\nc)$ covariance matrix $\P\equiv\expec{\x\x^t}$ as
$\P_{ij}=\sigma_i\sigma_j\R_{ij}$, where $\sigma_i\equiv\P_{ii}^{1/2}$
is the {\rms} fluctuation of the $\ith$ component and
$\R_{ij}\equiv\P_{ij}/\sigma_i\sigma_j$ is the dimensionless
correlation matrix.  The components $r_i\equiv R_{0i}$ clearly
correspond to the correlations we discussed earlier, except that there
is now one $r$ for each clan, a vector of correlations $\r$.  These
are the numbers we are interested in, although all we can measure
directly are the remaining degrees of freedom in $\R$, the
galaxy-galaxy correlation matrix $\Q$ (the $\nc\times\nc$ submatrix in
the lower right corner).  What constraints can we place? Consider
first the simpler case involving only $\nc=2$ two clans. Then
\beq{CorrMatrixEq}
\R
=\left(\bs\begin{tabular}{cc}
1	&$\r^t$\\
$\r$	&\Q\\
\end{tabular}\bs\right)
=\left(\bs\begin{tabular}{ccc}
1	&$r_1$	&$r_2$\\
$r_1$	&1	&$\rho$\\
$r_2$	&$\rho$	&1\\
\end{tabular}\bs\right),
\eeq
where $\rho$ denotes the one number that we can measure: the
correlation between clans 1 and 2.  Since a correlation matrix cannot
have negative eigenvalues, $\det\R\ge 0$. After some algebra,
this gives the constraint
\beq{rConstraintEq1}
|r_1-\rho r_2|\le\left[(1-\rho^2)(1-r_2^2)\right]^{1/2}.
\eeq
Thus the larger we make $r_2$, the more tightly constrained $r_1$
becomes, and vice versa. If $r_2=1$, then the right hand side
vanishes, forcing $r_1=\rho$.  This gives an upper bound on the
smaller of $r_1$ and $r_2$.  The most weakly constrained case is
clearly $r_1=r_2$, so we obtain
\beq{MinLimEq}
\hbox{min}\,\{r_1,r_2\}\le\left[{1+\rho\over 2}\right]^{1/2}.
\eeq
For the case with $\nc=4$ clans, we get analogous inequalities for
each of the $\nc(\nc+1)/2=10$ pairs of clans. Additional more
complicated constraints follow from requiring a nonnegative
determinant for all $4\times 4$-submatrices and for the whole $5\times
5$ matrix, although these will not be considered here.

\begin{center}
{\footnotesize
{\bf Table 1.} Correlations and relative 
bias for the four galaxy clans.
\smallskip

\noindent
\begin{tabular}{|cc|cccc|}
\hline
$i$	&$b_i/b_4$	&\multicolumn{4}{c|}{Galaxy correlation matrix $\Q$}\\
\hline	
1	&2.97$\pm.04$	&1		&.63$\pm$.02	&.44$\pm$.02	&.42$\pm$.03\\
2	&1.28$\pm.01$	&.63$\pm$.02	&1		&.88$\pm$.02	&.81$\pm$.02\\
3	&1.21$\pm.01$	&.44$\pm$.02	&.88$\pm$.02	&1		&.76$\pm$.03\\
4	&1.00$\pm.01$	&.42$\pm$.03	&.81$\pm$.02	&.76$\pm$.03	&1\\
\hline
\end{tabular}
}
\end{center}

\vskip2mm

\subsection{Measuring the galaxy correlation matrix}

$\Q_{ij}=\G_{ij}/(\G_{ii}\G_{jj})^{1/2}$, 
where $\G$ is the $\nc\times\nc$ galaxy-galaxy
covariance matrix (the bottom right submatrix of $\P$).
Based on the analysis of Tegmark {\etal} (1998, hereafter
T98), we compute an estimate $\G$ of the
form $\G_{ij} \equiv {\g^{(i)}}^t\E\g^{(j)}$ minus shot noise, removed
as in Appendix B of T98 by simply omitting self-pairs.  The pair
weighting is given by a matrix $\E\equiv\N^{-1}\F\N^{-1}$, where $\N$
is some typical shot noise covariance matrix -- since it is different
for each clan, we use the average $\N=\nc^{-1}\sum_i\N^{(i)}$ as a
compromise.  We make the choice $\F=\I$, which corresponds to
measuring the $\rms$ fluctuations in the cells.  Computing the window
function $W$, defined by $\expec{\G_{ij}} = \int \P_{ij}(k) W(k)
d^3k$, we find that $\G_{ij}$ mainly probes scales around
$\lambda\equiv 2\pi/k\sim 10 h^{-1}\Mpc$.

Converting $\G$ back to $\Q$, we calculate error bars for its
elements and all quantities derived below by generating $10^4$ Monte
Carlo realizations of Poisson shot noise for $N_\alpha^{(i)}$, adding
this to the observed values and processing the result in exactly the
same manner as the real data.  Since our primary goal is to rule out
$r_i=1$, we clearly do not care about sample variance in $\Q$: if
biasing were simple, one would measure $r_i=1$ in all samples.
When reading Tables 1 and 2, the reader should thus bear in mind that
the error bars reflect shot noise only.

The results are shown in Table 1. The $\second$ column confirms that
the clustering strength drops with clan number, as shown by Bromley
{\etal} (1998a).  All off-diagonal elements in the correlation matrix
are seen to be significantly below unity, ranging from the $88\%$
correlation between clans 2 and 3 to a mere $42\%$ correlation between
the earliest and latest galaxy types. Correlations generally decrease
with separation in clan number, as expected.  Applying
equations\eqn{rConstraintEq1} or\eqn{MinLimEq} to this table gives a
plethora
of constraints on $r_i$.
For example, \eq{MinLimEq} tells us that either $r_1$ or $r_4$ 
must be below $\sqrt{(1+0.42)/2}\approx 84\%$.

\subsection{The matter as a principal component}

Let $\lambda_1\ge\lambda_2\ge...\ge\lambda_{n_c}$ be the 
eigenvalues of $\G$, sorted from largest to smallest, 
with $\e_i$ the
corresponding unit eigenvectors ($\G\e_i=\lambda_i\e_i$,
$\e_i\cdot\e_j=\delta_{ij}$).  It is instructive to decompose the
fluctuation vector $\x$ into its uncorrelated principal components
$y_i\equiv\e_i\cdot\x$:
\beq{PCAeq}
\x=\sum_{i=1}^{\nc} \e_i y_i,
\eeq
where $\expec{y_i y_j}=\delta_{ij}\lambda_i$.
If biasing were simple and linear, we would have
$\x=\b\,\delta$ for some vector $\b$ of bias values $b_i$.
\Eq{PCAeq} would therefore have
only one term, with $\e_1\propto\b$ and $y_1\propto\delta$.
Additional terms in \eq{PCAeq} would correspond to random physical
processes, uncorrelated with $\delta$, that are pushing $r_j$ below
unity.  If we want a model with a minimal amount of stochasticity, we
should therefore interpret the first (largest) principal component as
tracing the underlying matter distribution, \ie, assume that
$\delta=ay_1=a\e_1^t\x$ for some constant $a$. Since
$\expec{\x\x^t}=\G$, this assumption gives
$\expec{\x\delta}=a\expec{\x\x^t}\e_1=a\G\e_1=a\lambda_1\e_1$,
$\expec{\delta^2}=a^2\e_1^t\expec{\x\x^t}\e_1=a^2\e_1^t\G\e_1=a^2\lambda_1$,
and hence the correlation coefficients
\beq{Minimal_r_Eq}
r_j = 
{\expec{x_j\delta}\over\left[\expec{x_j}^2\expec{\delta}^2\right]^{1/2}}
= \sqrt{\lambda_1\over\G_{jj}}(\e_1)_j.
\eeq
These coefficients are shown in the last row of Table 2.  The fraction
of the variance of clan $j$ that is caused by matter fluctuations is
$\expec{(\e_1)_j^2y_1^2}/\expec{x_j^2} =
(\e_1)_j^2\lambda_1/\G_{jj}=r_j^2$, simply the square of the
correlation coefficient, so this is a useful way to interpret $r_j$.

Table 2 is indeed consistent with the hypothesis that the first
principal component traces the matter: since all of its four
components (boldface) have the same sign, we can interpret them as the
(relative) bias factors that the clans would have if there were no
randomness ($\lambda_2=\lambda_3=\lambda_4=0$).  Since
$(\e_1)_j\propto\G_{jj}^{1/2}\propto b_j r_j$, we can also interpret
them as the best fit slopes in linear regressions of $x_j$ against
$\delta$, the quantity that Dekel \& Lahav (1998) call simply ``$b$''.
The sharp decrease of $(\e_1)_j$ towards later clan types (their
ratios are $5.1:1.7:1.3:1$) is thus caused by the joint decrease of
both the {\it measured} relative bias $b_j$ (the $\second$ column of
Table 1) and the correlation $r_j$ .

\begin{center}
{\footnotesize
{\bf Table 2.} Principal components decomposition of the fluctuations.
\smallskip
\noindent
\begin{tabular}{|c|c|rrrr|}
\hline
i	&$\lambda_i$	&\multicolumn{4}{c|}{Components of eigenvector $\e_i$}\\
\hline	
1	&.564$\pm$.013	&{\bf.91$\pm$.01}&{\bf.31$\pm$.01}&{\bf.23$\pm$.01}&{\bf.18$\pm$.01}\\
2	&.122$\pm.004$	&-.41$\pm$.01	&.51$\pm$.01	&.60$\pm$.01	&.46$\pm$.02\\
3	&.016$\pm.002$	&.01$\pm$.02	&-.08$\pm$.13	&-.56$\pm$.16	&.82$\pm$.15\\
4	&.008$\pm.001$	&-.08$\pm$.01	&.80$\pm$.03	&-.52$\pm$.10	&-0.28$\pm$.14\\
\hline
\multicolumn{2}{|c|}{Clan correlation $r_j$}&.98$\pm$.002	&.77$\pm$.02	&.60$\pm$.02	&.57$\pm$.03\\
\hline
\end{tabular}
}
\end{center}

\section{Conclusions}


Our basic conclusion is that bias is complicated. Rather than being
describable by a single constant $b_i$ for the $\ith$ galaxy type,
evidence is mounting that bias is
\begin{enumerate}
\itemsep0cm
\item stochastic and/or nonlinear, requiring a $\second$ quantity $r_i$ to 
characterize $\second$ moments,
\item scale dependent ($b_i$ and $r_i$ depend on $k$),
\item time-dependent ($b_i$ and $r_i$ depend on redshift).
\end{enumerate}
Complication (1) was predicted by Dekel \& Lahav (1998), and we have
observationally confirmed it here.  Complication (2) has long been
observed on small scales (see {\eg} Mann {\etal} 1996, B98 and
references therein), while Complication (3) was predicted by Fry
(1996) and TP98 and is gathering support from both simulations (Katz
{\etal} 1998; B98) and observations (Giavalisco {\etal} 1998).

However, this is not cause for despair.  The scale dependence is
predicted to abate on large scales (Scherrer \& Weinberg 1998), and
the time-evolution due to gravity is calculable (TP98).  As long as we
limit ourselves to second moments (power spectra {\etc}),
stochasticity merely augments $P_i(k)$ and $b_i(k)$ with an additional
function $r_i(k)$.  Furthermore, our constraints on $r_i$ using galaxy
data alone --- without knowledge the underlying mass distribution ---
suggest strong regularities: The more similar two morphological types
are, the stronger they are correlated.

More strikingly, our tentative identification of the $\first$
principal component of the galaxy covariance matrix as the underlying
matter distribution is in excellent agreement with the recent
simulations of B98: galaxies with high bias are almost perfectly
correlated with matter (we find $r_1\sim 98\%$), whereas the less
biased populations have weaker correlations (we find $r\sim
60\%-80\%$). Thus matter clustering explains only
$r^2\sim 40\%-60\%$ of the variance of these 
galaxies.

This analysis is merely a first step towards observationally measuring the
parameters of stochastic biasing.  It remains to explore the
scale-dependence and time-evolution of $r$ as well as probing
higher-order moments of the joint distribution of galaxy types (e.g.,
Lahav \& Saslaw 1992).  However, our success in constraining $r$ in
the absence of difficult mass measurements is an encouraging
indication for the analyses of upcoming galaxy surveys: With the
techniques presented here, along with theoretical and numerical
modeling, it may be possible to understand bias well enough to
realize the full potential of these surveys for measuring fundamental
cosmological parameters.


\smallskip
We thank the LCRS team for kindly making their data public and John
Bahcall, Michael Blanton, Avishai Dekel, Ofer Lahav, Jim Peebles and
Uro\v s Seljak for useful discussions. MT was funded by NASA
though grant NAG5-6034 and Hubble Fellowship HF-01084.01-96A from
STScI, operated by AURA, {\frenchspacing Inc.} under NASA contract
NAS5-26555.  BB was funded by NSF Grant PHY 95-07695 while
at the Harvard-Smithsonian CfA and acknowledges
computing support from NASA and the Caltech Center for Advanced
Computing Research.




\end{document}